\documentclass[12pt,preprint]{aastex}

\def\simlt{\lower.5ex\hbox{$\; \buildrel < \over \sim \;$}}
\def\simgt{\lower.5ex\hbox{$\; \buildrel > \over \sim \;$}}
\def\d10{d_{\rm 10}}
\def\kms{km s$^{-1}$}
\def\schi{{\sc Hi}\ }
\def\msun{{$M_\odot$}}

\def\alphaJ{{\alpha_{2000}}}
\def\deltaJ{{\delta_{2000}}}

\def\vlsr{v_{\rm LSR}}

\def\vexp{v_{\rm exp}}

\def\fvw{FVW 190.2+1.1}

\shorttitle{A ``MISSING'' SUPERNOVA REMNANT}
\shortauthors{Koo et al.}

\begin{document}


\title{A ``MISSING'' SUPERNOVA REMNANT REVEALED BY THE
21-CM LINE OF ATOMIC HYDROGEN}

\author{Bon-Chul Koo and Ji-hyun Kang}
\affil{Department of Physics and Astronomy, Seoul National University, Seoul 151-742, Korea}
\email{koo@astrohi.snu.ac.kr; kjh@astro.snu.ac.kr}

\and
\author{C. J. Salter}
\affil{National Astronomy and Ionosphere Center, Arecibo Observatory, HC\,3 Box 53995, 
Arecibo, PR~00612, USA} 
\email{csalter@naic.edu}

\begin{abstract}
Although some 20--30,000 supernova remnants (SNRs) are expected to exist in
the Milky Way,
only about 230 are presently known. This
implies that most SNRs are ``missing''. Recently, we proposed that
small ($\simlt 1^\circ$), faint, high-velocity features seen in
large-scale 21-cm line surveys of atomic hydrogen ({\sc Hi}) in the
Galactic plane could be examples of such {\it missing} old SNRs.  Here
we report on high-resolution \schi observations of one such candidate,
FVW 190.2+1.1, which is revealed to be a rapidly expanding ($\sim
80$~\kms) shell.  The parameters of this shell seem only consistent
with FVW 190.2+1.1 being the remnant of a SN explosion that occurred in
the outermost fringes of the Galaxy some $\sim 3\times 10^5$~yr ago.
This shell is not seen in any other wave band suggesting that it
represents the oldest type of SNR, that which is essentially invisible
except via its \schi line emission. 
FVW 190.2+1.1 is one of a hundred 
``forbidden-velocity wings" (FVWs) recently identified in the Galactic plane,
and our discovery suggests that many of these are likely to be among
the oldest SNRs. We discuss the possible link between FVWs and
fast-moving atomic clouds in the Galaxy.

\end{abstract}


\section{Introduction}

From radio and X-ray studies of the Milky Way, about 230 supernova
remnants (SNRs) are known \citep{gre04}. However, a supernova (SN)
outburst occurs every 30--50 yr on average in a galaxy like the Milky
Way \citep{van94,cap99}.  Hence, if $\sim 10^6$~yr is taken to be 
the life time of a SNR, the total number of remnants in our Galaxy 
should be (2--3)$\times 10^4$. This suggests that most SNRs are 
``missing''. 

Recently, we proposed that small ($\simlt 1^\circ$), faint, 
high-velocity features seen in
large-scale 21-cm line surveys of atomic hydrogen ({\sc Hi}) in the
Galactic plane could be examples of such {\it missing} old SNRs
\citep{koo04, kan04}. These ``forbidden-velocity wings''  (FVW) 
appear in longitude-velocity diagrams as low-level, yet significant,
bumps protruding from the regular Galactic \schi emission (Fig. 1).  
The smooth boundary of the \schi distribution in Fig. 1 is 
given by the extent and rotational properties of the Galactic gaseous disk.
Small-scale features lying significantly beyond these boundaries
are likely to have originated from energetic phenomena. 
The strong, discrete emission feature at 
$l=196^\circ$ is a good example, and is associated with 
the collision of high-velocity clouds (HVCs) with the Galactic plane
\citep{tam97}.
FVWs are distinguished from the HVCs
in that they appear as protrusions from the Galactic background emission, 
and not as separate discrete peaks. In Fig. 1, 
FVW 190.2+1.1, extending to $\vlsr\simeq
+80$~\kms\ at $l\approx 190^\circ$, is a typical example.

We noted that there are many such 
FVWs of unknown origin in the Galactic
plane and that these look similar to \schi features observed towards
old SNRs.  In the late stages of their evolution,
SNRs are expected to develop dense atomic shells which emit in the
\schi 21-cm line.  However, because of confusion with emission from
atomic gas along the line-of-sight, the \schi radiation from an SNR
shell is only observable when its expansion velocity extends
significantly beyond the maximum and/or minimum velocities permitted by
Galactic rotation, i.e., when its \schi emission appears as a
protrusion in velocity (see Fig. 1).  Systematic studies have been
previously made towards two hundred {\it known} SNRs, but 
high-velocity \schi wings have {\it only} been detected towards two
dozen \citep{koo91, koo04b}. It was pointed out that this relatively
low detection rate is because these studies focused exclusively on
cataloged radio-continuum/X-ray emitting SNRs, hence excluding old
SNRs which have become too faint to be recognized in these ways
\citep{koo04}.  Recent sensitive, high-resolution, radio continuum
and X-ray observations have discovered several very faint SNRs
indicating that there are indeed many old SNRs awaiting discovery
\citep{bro04, sch02}.  However, while we have proposed that the FVWs
are candidates for such {\it missing} old SNRs, existing  
Galactic plane \schi surveys have insufficient resolution and/or 
sensitivity to reveal their true natures \citep{har97, mcc05, tay03}. 

In this Letter, we present the results of high-resolution \schi
observations of FVW 190.2+1.1, the FVW shown in Fig. 1, which show it
to be a rapidly expanding shell, probably originating from a SN
explosion.

\section {Observations}

In February 2004 and 2005, we used the Arecibo 305-m telescope\footnote{
The Arecibo Observatory is
part of the National Astronomy and Ionosphere Center, which is operated by
Cornell University under a
cooperative agreement with the National Science Foundation.}
to obtain a high-resolution (${\rm HPBW} = 3.\!'4$) \schi image of FVW
190.2+1.1. Total-intensity 21-cm spectra with a total bandwidth of
3.125~MHz and 1024 frequency channels were obtained using
the dual-channel, linear polarization, L-wide receiver.
This gave a velocity coverage of 660~\kms\ and 
a velocity resolution of 1.29~\kms\ after Hanning smoothing. 
A rectangular area of $2.\!^\circ 0 \times 1.\!^\circ 6$
centered at $(\alpha_{2000},\delta_{2000})=(6^{\rm h} 12^{\rm m} 00.\!^{\rm s}0, 
20^\circ 32' 00'')$ was mapped using fixed-azimuth drift scanning,
with a $1.\!'7$ step in declination. The resulting spectra were converted 
into antenna temperature, and then gridded via convolution
to produce a data cube with a pixel size of $1.\!'7$ and a final
HPBW of $3.\!'9$. A polynomial baseline of order 2--5 was subtracted from
each spectrum. The rms ($1\sigma$) noise in the cube is $\sim 0.05$~K.

Additionally, in January 2005 we searched for 2.5-GHz synchrotron emission from the
shell of FVW 190.2+1.1 using the Green Bank Telescope (GBT).  Both
linear polarizations were recorded using the dual
polarization S-band receiver centered at 2.53 GHz and the GBT digital
continuum receiver with a bandwidth of 12.5 MHz.  A
rectangular area of $2.\!^\circ 5 \times 2.\!^\circ 0$ centered at
$(\alpha_{2000},\delta_{2000})=(6^{\rm h} 13^{\rm m} 00.\!^{\rm s}0,
20^\circ 40' 00'')$ was mapped using ``On-the-fly" data acquisition, 
with a
scanning rate of $2^\circ/{\rm minute}$ along right ascension and a
$2.\!'0$ step in declination. The sampling interval was 0.1 s.
Noise power was injected on alternate samples to permit
calibration of the total signal, (including emission from the
atmosphere, ground, and the Galactic background), into system
temperature.
We obtained data for just the ``source component" 
by subtracting a smooth baseline from each scan.  The resulting data
were gridded by convolution to produce an image with a pixel size of
$2.\!'0$ and a HPBW of $4.\!'6$. We have compared the surface
brightness across the intense, extended radio source Sh 2-252, 
situated within the
field, to that from the Bonn 11-cm survey \citep{fur90}.  There was an
excellent correlation between the two, although the brightness scale of 
the present data was
lower than that of the Bonn image by 23\% and 37\% for our
respective orthogonal
polarizations.  We have scaled our data to that from Bonn by
multiplying each polarization by the appropriate constant value.  The final
image was produced by combining the brightness distributions for the two
polarizations, and smoothing this with a Gaussian beam of
one-pixel dispersion, giving a HPBW of $4.\!'7$.  The rms ($1\sigma$)
noise on the image is $\sim 6$~mK.

\section {Results}

Fig. 2 shows the Arecibo HI images of FVW 190.2+1.1 for different velocity intervals, 
together with a three-color map generated from these.
An $\sim 1^\circ$-sized, mildly elliptical, shell structure is
clearly seen, with its major axis aligned roughly east-west. 
The emission ring becomes smaller the more positive the velocity
(Fig. 3)
indicating that we are seeing that portion of a shell which is
expanding away from us.  
While the shell structure is visible
for velocities between about +25 and +84 \kms, 
the end-cap is not seen. 
The end-cap is expected to be faint because of the spreading of the
\schi intensity over a range of velocities due to 
turbulence within the shell
\citep{koo95, caz05}. However, the upper limit on
the \schi column density of the end cap of FVW 190.2+1.1 is
very low, e.g., $1.6\times 10^{18}$ cm$^{-2}$ ($3\sigma$) over velocities
between +85 and +95~\kms, suggesting that the shell is 
probably incomplete towards this direction.
The constant angular diameter at lower velocities indicates that
this should be close to the size
of the expanding shell, and that 
its systemic velocity is $v_0 \simlt +30$~\kms.  Another
pointer to $v_0$ exists. To the east of the region studied, for
$\vlsr=+35$--55~\kms, there is faint background emission whose intensity
drops steeply across the eastern ridge of the shell 
(see the G and B images in Fig. 2), 
indicating that this external emission is from ambient gas interacting
with the shell. 
We see only the wing portion of this emission and, 
by least-squares fitting of the 21-cm line
profile with several Gaussian components, we obtain a central velocity of
$\simgt +10$~\kms\ for this component. 
Therefore, we have $+10\simlt v_0 \simlt +30$~\kms\ and 
adopt $v_0=+20\pm 5$~\kms, 
where the error is a probable error \citep{bev69}.
In this direction, the LSR velocity is a shallow
function of distance so that the systemic velocity does not yield a 
good kinematic distance. Nevertheless, the systemic 
velocity is reliable, even if the distance to FVW 190.2+1.1 is only
poorly constrained. 

Expansion parameters for FVW 190.2+1.1 can be derived from the
velocity-size dependence shown in Fig. 3.
The simplest model is a spherical shell expanding radially with 
uniform velocity. For this, the dotted curve shows the best-fit model
which has an angular radius of $\bar a=0.\!^\circ 63$ and an
expansion velocity of $\vexp=77$~\kms. 
In practice, the velocity-size relation derived appears 
to possess more of a linear form than having the concave 
shape predicted by the model. 
This difference can be explained by turbulent motions within the 
shell, e.g. the \schi emission lines are broadened by turbulent motions,
and this broadening makes the relation appear more linear. For example, 
the solid curve in Fig. 3 shows the result for a partially complete 
spherical shell with the above expansion velocity, but slightly (3\%) larger 
diameter and a line-of-sight turbulent velocity dispersion of $10$~\kms.
We therefore consider that the simplest model provides
an acceptable description of the expansion properties of FVW 190.2+1.1, and 
adopt $\bar a=0.\!^\circ 63\pm 0.\!^\circ 03$ and an
expansion velocity of $\vexp=77\pm 6$~\kms. 
For convenience, we normalize the physical parameters to a distance of 10 kpc, 
giving a
geometrical-mean radius $R_s = 110d_{10}$ pc, while the major and minor
axes of the shell ($42'\times 34'$) correspond to $(120 \times
99)\d10$~pc, where $d_{10}$ is the distance to FVW 190.2+1.1 in units
of 10 kpc.  

We derive the mass of the shell using the channel maps. In each map, we use nested
elliptical rings to estimate how the mean \schi intensity varies with
distance from the geometrical center.  The elliptical rings have an
axial ratio of 1.24 ($=42'/34'$) and a thickness of $1.'7$ (=pixel
size) along the major and minor axes. 
The plot of the mean \schi intensity as a function of
radial distance reveals a bump which peaks at the location of the shell.  We fit
a smooth baseline and obtain the integrated \schi flux density in the
bump which can be easily converted to mass given that the \schi
emission is optically thin.  The derived \schi shell mass over the
velocity range of +40--80~\kms\ is $(1.4 \pm 0.3) \times 10^3
\d10^2$~\msun.  The mass per unit velocity interval increases as the
velocity decreases.  This is a general trend for rapidly expanding
shells and has been attributed to the clumpy nature of the interstellar
medium \citep[e.g.,][]{gio79, koo91}. In order to derive the total \schi
mass of the expanding shell, we need to estimate the mass in the
unobserved portion of the shell lying at lower velocities ($\simlt
+40$~\kms) where the background \schi emission dominates.  The
 mass distribution can be described by a Gaussian, and we
estimate the total mass by fitting a Gaussian centered at $v_0$. The
extrapolated total \schi mass of the shell is $6.5\times 10^3
\d10^2$~\msun.  Including the cosmic abundance of helium, the
corresponding kinetic energy of the shell is $E_K\simeq 5.4 \times
10^{50} \d10^2$~erg. The large extrapolation factor means that the
estimated total mass has a greater relative uncertainty. However, if the
energetic phenomenon that produced the expanding shell had spherical
symmetry, the derived kinetic energy should be reasonably ($\sim 30$\%)
accurate.

As seen from Fig. 2 (lower right panel), no 2.5~GHz-continuum emission is 
detected from \fvw\ with an ($3\sigma$) upper limit of $0.02$ K,
corresponding to a 1-GHz surface brightness of $7\times 10^{-23}$
W~m$^{-2}$~Hz$^{-1}$~sr$^{-1}$,  
assuming that flux density varies with frequency 
as $\nu^{-\alpha}$, with $\alpha=0.5$.
The shell is seen in neither the
IRAS maps at 60 and 100 $\mu$m nor the distribution of their ratio.  It
is also not seen in DSS optical, VTSS H$\alpha$, or ROSAT
All-sky X-ray survey images \citep{fin03, ros00}.

\section {Discussion}

FVW 190.2+1.1 has a large ($\sim 80$~\kms) expansion velocity. 
Although there have been discoveries of expanding 
\schi shells and supershells, 
their expansion velocities are usually $\simlt 20$~\kms\ 
\citep{sti01, uya02, hei79}.
\schi shells with expansion velocities comparable to that of FVW 190.2+1.1
have only been found toward old SNRs \citep[][and references therein]{gio79, koo91, koo04}, 
suggesting a SN origin for this shell.  
For a SNR shell, the
initial explosion energy, $E_{\rm SN}$, can be estimated from
$E_{\rm SN}=6.8\times 10^{43} n_0^{1.16} R_s^{3.16} \vexp^{1.35}
\zeta_m^{0.161}$ erg, where $n_0$ is the ambient density of hydrogen nuclei
in cm$^{-3}$, $R_s$ is in pc, $\vexp$ is in \kms, and $\zeta_m$ is the
metallicity \citep{cio88}.  We estimate the ambient density for FVW~190.2+1.1 by 
assuming that
the current \schi mass in the shell was initially distributed uniformly 
over the volume contained within the shell. 
This yields $n_0=0.048 \d10^{-1}$~cm$^{-3}$.
The metallicity in the outer Galaxy is significantly lower than in the
solar neighborhood, and we adopt $\zeta_m=0.2$ which is the metallicity
at $d\sim 10$ kpc based on \citet{mac99}. However, since the SN energy depends 
only very weakly on the metallicity, the following discussion will not
be affected by our adopted metallicity.
Substituting
these values, $E_{SN}=1.5 \times 10^{51} \d10^2$~erg. 
For the canonical value of $E_{SN}=1\times 10^{51}$~erg, this implies
a distance of 8~kpc, corresponding to $R_s=88$~pc and a dynamical age of
$t\simeq 0.3 R_s/v_s\simeq 3.4\times 10^5$~yr. The distance appears rather
larger than expected for a SN explosion in the Galaxy, but is not unreasonable.
It is also consistent with the empirical surface brightness-diameter
($\Sigma-D$) relation for SNRs. 
Extrapolating the $\Sigma-D$ relation of \cite{cas98},
we note that our 1-GHz continuum brightness upper limit 
is comparable to that predicted for a SNR shell of radius, $R_s \sim 100$~pc. 
Despite the large scatter of the
$\Sigma-D$ relation, this is also consistent with FVW 190.2+1.1 being a
large, old SNR. We further note that several Sharpless HII
regions have been discovered at such distances toward the anticenter
region, including Sh 2-259 which is at 8.3 kpc toward $(l,b)=(192.^\circ 91,
0.^\circ 63)$ and has a similar radial velocity to FVW 190.2+1.1
($22.8\pm 0.5$~\kms) \citep{mof79, bra93}. \cite{mof79} proposed
the existence of
a remote spiral arm connecting these HII regions from $l=150^\circ$ to
$220^\circ$, and it is quite possible that FVW 190.2+1.1 is a SNR  
associated with this.

Large,
fast-expanding shells  can also be produced by stellar winds from OB
stars. The wind luminosity required to drive FVW 190.2+1.1 would
be $L_w =
(77\,E_K) /(9\,R_s /\vexp) = 8.3 \times 10^{37} \d10$ erg s$^{-1}$
\citep{wea77}, which even for a distance as close as 
1 kpc corresponds to a star earlier than O6V
\citep{abb82}.  The only early-type star with a comparable luminosity
within the field of Fig. 2 is HD 42088, the O6.5V star exciting the
Sharpless HII region Sh 2-252 at 2 kpc \citep{car95}.  However, this star is
situated outside of the \schi shell, and no morphological connection is
apparent between the HII region and the shell. Also, if the shell
were to have been produced by an O6.5 star, 
it would not be expected to be neutral
because of the strong ionizing radiation.
We further note that many high velocity clouds (HVCs) forming a
complex chain have been observed near $-200$~\kms\ in this region of
the anticenter \citep{mir90}.  These HVCs are colliding with the Galactic
gaseous disk and generating a huge ($\sim 30^\circ$) \schi
supershell, essentially all of whose gas is moving at
negative radial velocities \citep{tam97}.  FVW 190.2+1.1 is much smaller, and is seen at
positive radial velocities, making it unlikely to be associated with the
HVCs.  
Summarizing, the lack of a viable alternative explanation for
its characteristics, strongly supports FVW190.2+1.1 being an old SNR.  
This indicates that a SNR expanding
as fast as 80~\kms\ can be ``dark'', i.e. essentially invisible in
the radio continuum.

FVW 190.2+1.1 is one of a hundred FV wings recently identified in the
Galactic plane \citep{kan04}. 
These are found in both the inner and outer
Galaxy, and also rather far from the Galactic plane ($|b|\simlt 12^\circ$). 
Comparison with existing catalogs has shown that only
$14$\% of these FVWs are associated with known external galaxies, HVCs,
or SNRs, and the natures of most remain a puzzle. Nevertheless, our
present discovery suggests that many 
are likely to be among the oldest SNRs.
We note that very recent Arecibo \schi observations 
made by ourselves of about ten FVWs do reveal 
expanding shell structures towards half of them, which supports this conjecture. 
The  other half show arc-, or ring-like structures which could also
be fragments of disrupted old SNR shells. 
It has been known for some time that our Galaxy 
contains \schi gas with a large velocity dispersion and extending far from 
the disk -- the \schi halo \citep{kul85, loc02}.
Recent high-resolution observations reveal 
small ($\sim 10$~pc), fast-moving clouds that constitute this high
velocity 
dispersion gas \citep{loc02, sti05}. 
Thus, it was suggested that our Galaxy may possess a population of fast-moving clouds 
distributed both throughout the disk and up into the halo, and that 
about one half of 
the Galactic \schi halo could be in the form of discrete clouds
\citep{loc02, sti05}.
These fast-moving clouds have similar properties to FVWs in
that both appear at forbidden velocities and, for both, 
the higher the peculiar velocity, the smaller the mass. 
However, the fast-moving clouds generally have lower velocities
than the FVWs, are of higher \schi column density, and 
appear as distinct clouds. It is possible that these clouds are 
also fragments of old SNR shells, but of older, more disrupted examples than 
those associated with FVWs. 
The nature of FVWs and their relation to 
the Galactic \schi structure should soon be revealed by new high-sensitivity,
high-resolution \schi surveys such as those recently undertaken
with the Arecibo L-band Feed Array (ALFA).

\acknowledgements

We thank Phil Perillat for his help with data reduction at Arecibo. 
We also wish to thank the Green Bank staff for their support, particularly 
Tony Minter, Jay Lockman, Dana Balser, and Carl Bignell.
We would like to thank Chris McKee for helpful comments on the original 
manuscript. We also thank the anonymous referee for comments which 
improved the presentation of this paper. 
This work was supported by the Korea Science 
and Engineering Foundation (ABRL 3345-20031017). J.-h. K. has been supported in part
by the BK 21 program.

\clearpage

\begin{figure}
\epsscale{.80}
\plotone{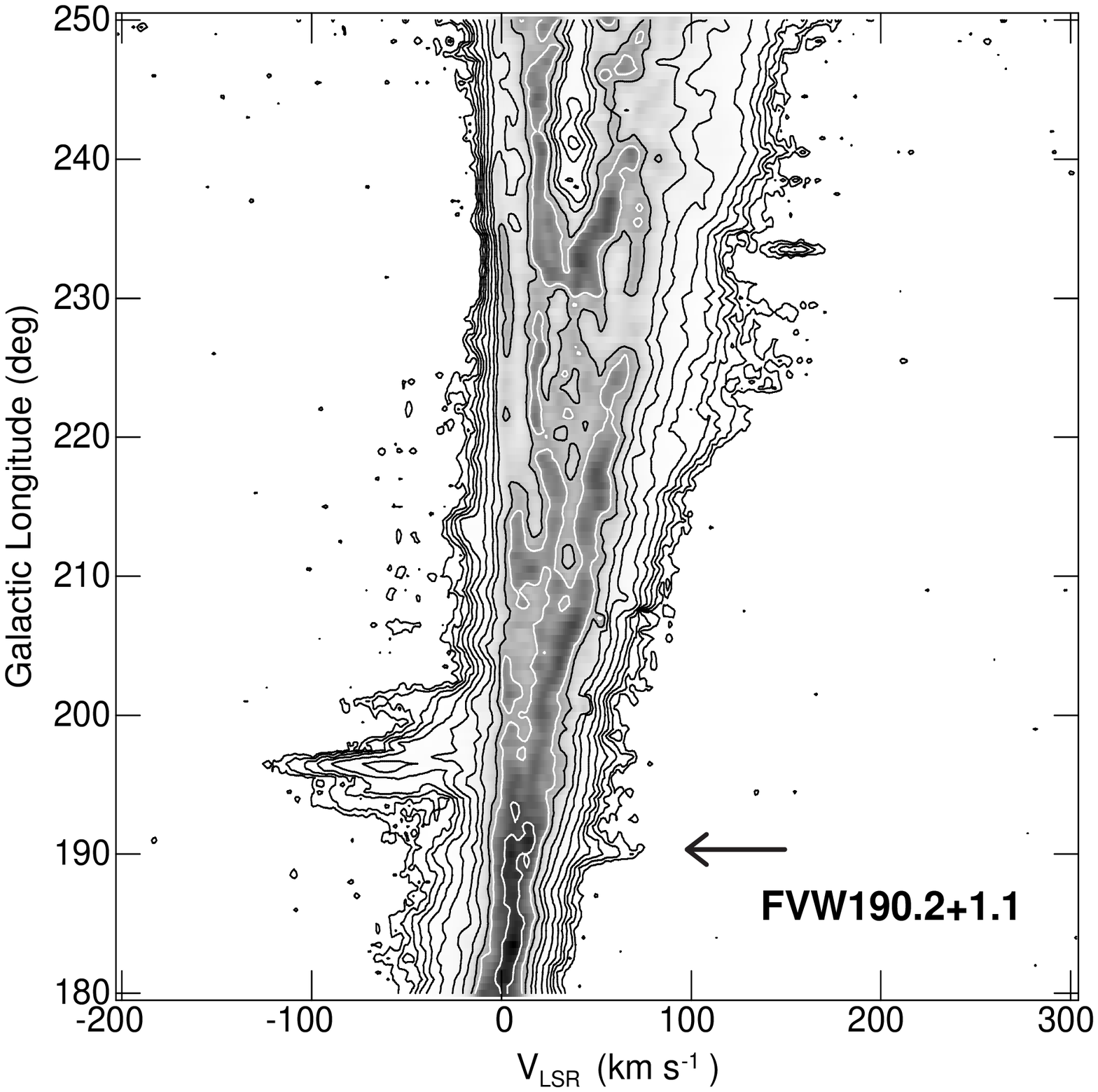}
\caption{A large-scale longitude-velocity map of atomic hydrogen emission 
at a Galactic latitude of +$1.\!^\circ 0$  
made using data from the Leiden-Dwingeloo survey at 
an angular resolution of $36'$ \citep{har97}.
The location of the forbidden-velocity wing FVW 190.2+1.1 is marked.
The contour levels are 0.15, 0.25, 0.4, 0.6, 1, 2, 5, 10, 30, 50, and 100 K
in brightness temperature.
}
\end{figure}
\clearpage

\begin{figure}
\epsscale{1.0}
\plotone{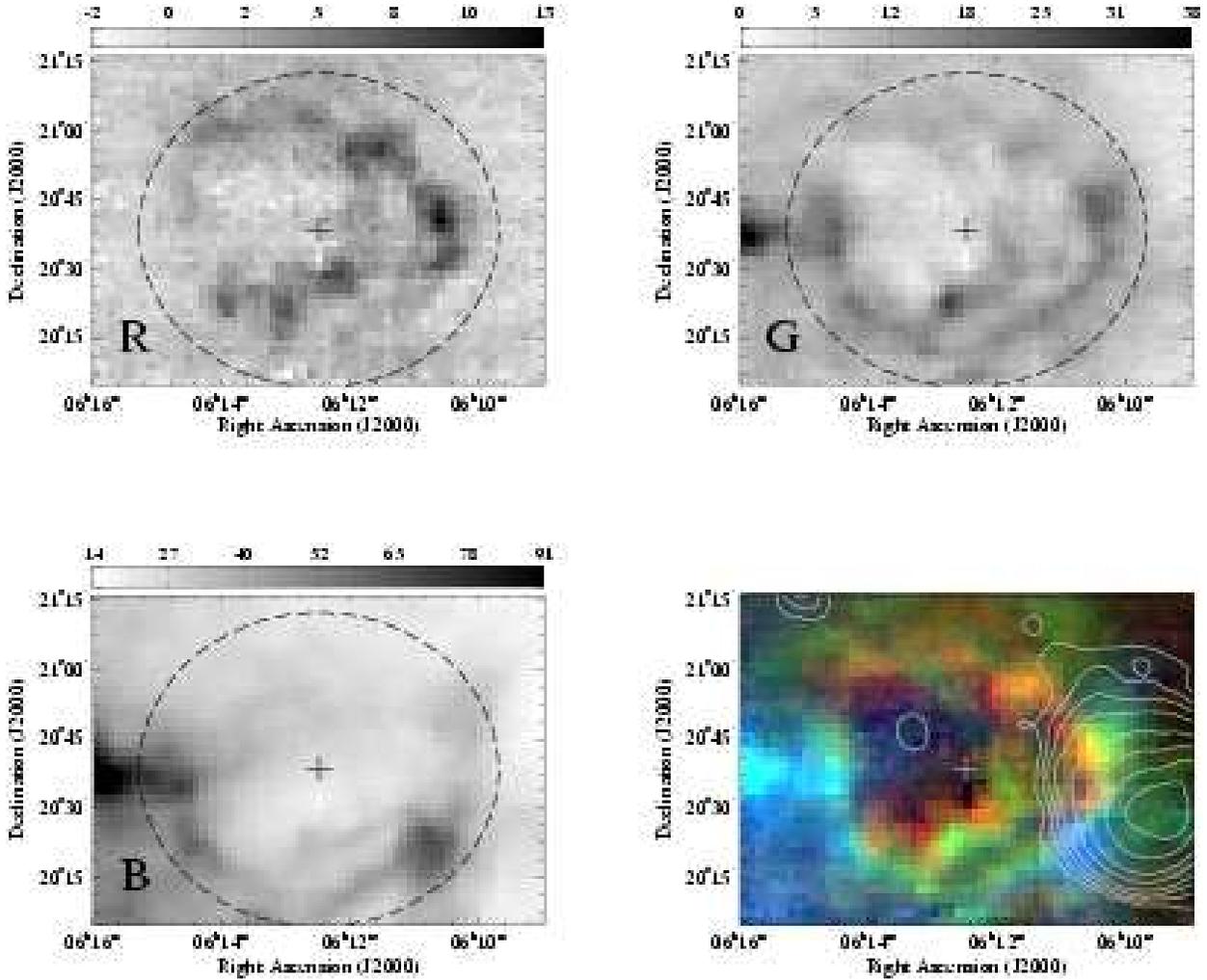}
\caption{The Arecibo \schi emission images 
for FVW 190.2+1.1 integrated over LSR velocities 
+59--79~\kms\ (R), +41--59~\kms\ (G), and +31--41~\kms\ (B). 
Grey scales (in K \kms) are given above individual images. 
The lower right panel is  a three-color image generated from these.
Note the well-defined 
shell structure, while the varying size with velocity indicates that 
we are looking at the 
receding portion of an expanding shell. 
The dashed ellipse marks the derived size of the expanding shell.
The geometrical center of the shell (marked by a cross) is 
$(\alphaJ,\deltaJ)=(6^{\rm h} 12^{\rm m} 29^{\rm s}, +20^\circ 38'\!.4)$ or 
$(l,b)=(190.\!^\circ 23,1.\!^\circ 14)$  with an uncertainty of $1.\!'2$. 
The contours superposed on the lower right image show the 
distribution of 2.5-GHz radio-continuum intensity obtained using the 
Green Bank Telescope; a smooth background emission has been subtracted. 
The contour levels are 0.02, 0.05, 0.1, 0.2, 0.5, 1, 2, and 3 K
in brightness temperature.
The bright radio source near the western boundary of
the map is Sh 2-252, an HII region which is not considered to 
be associated with FVW 190.2+1.1.
}
\end{figure}
\clearpage

\begin{figure}
\epsscale{.80}
\plotone{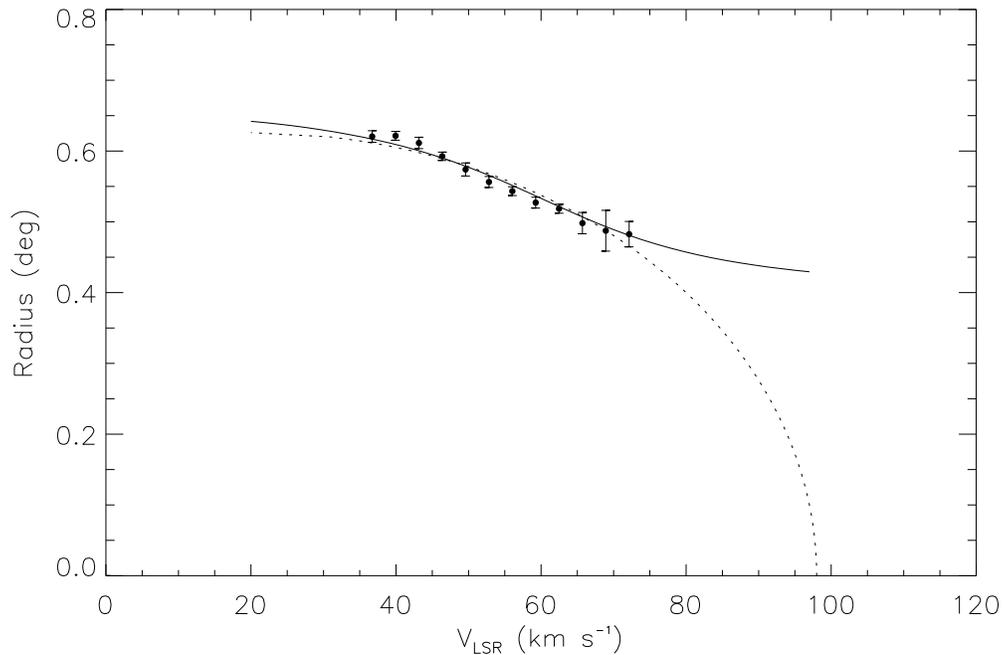}
\caption{Variation of geometrical-mean radius $\bar a$ with LSR velocity
for FVW 190.2+1.1.
We assume that the shell is axisymmetric, with the axis of symmetry
lying along the line of sight, so that $\bar a$ becomes simply its
radius of 
circular cross section. The error bars represent 
$1\sigma$ uncertainties.
The dotted curve is the best-fit profile for a spherical 
shell expanding radially with uniform velocity. 
The solid curve shows the result for a partially complete 
spherical shell with the same expansion velocity, but slightly (3\%) larger
radius, and 
with a line-of-sight turbulent velocity dispersion of $10$~\kms.
}
\end{figure}
\clearpage

\end{document}